\documentclass[aps, prc, superscriptaddress, nofootinbib, preprintnumbers,twocolumn]{revtex4}
\usepackage{amsmath}
\usepackage{graphicx}
\usepackage{color}

\newcommand{\chushi}[1]

\begin{document}
 \title{{\bf  Charged pions tagged with 
 polarized photons probing  
 strong CP violation \\
 in a chiral-imbalance medium}
 \vspace{5mm}}
\author{Mamiya Kawaguchi\footnote{mkawaguchi@hken.phys.nagoya-u.ac.jp}}
\author{Masayasu Harada\footnote{harada@hken.phys.nagoya-u.ac.jp}}
      \affiliation{ Department of Physics, Nagoya University, Nagoya 464-8602, Japan.}    
\author{Shinya Matsuzaki\footnote{synya@hken.phys.nagoya-u.ac.jp}}   
      \affiliation{ Department of Physics, Nagoya University, Nagoya 464-8602, Japan.}    
\affiliation{ Institute for Advanced Research, Nagoya University, Nagoya 464-8602, Japan.}
\author{Ruiwen Ouyang\footnote{ruiwen.ouyang@gmail.com}}
   \affiliation{ College of Physics, Jilin University, Changchun, 130012, China}
\date{\today}

\begin{abstract}
It is expected 
that in a hot QCD system, 
a local parity-odd domain can be produced due to  
nonzero chirality,   
which is induced from the difference of winding numbers carried by 
the gluon topological configuration (QCD sphaleron). 
This local domain is called the chiral-imbalance medium 
characterized by nonzero chiral chemical potential,   
which can be interpreted as the time variation of the strong CP phase.  
We find that the chiral chemical potential 
generates the parity breaking term in 
the electromagnetic form factor of charged pions. 
Heavy ion collision experiments could observe 
the phenomenological consequence of this parity-odd form factor 
through the elastic scattering of a pion and a photon 
in the medium.   
Then we quantify the asymmetry rate of the parity violation  
by measuring the polarization of 
the photon associated with the pion, 
and discuss 
how it could be measured in a definite laboratory frame. 
We roughly estimate the typical size of the asymmetry, 
just by picking up the pion resonant process,  
and find that the signal can be sufficiently larger than 
possible background events from 
parity-breaking electroweak process.  
Our findings might provide a novel possibility to make a 
manifest detection for the remnant of the strong CP violation.  
\end{abstract}
\maketitle

\section{Introduction}

It is still mysterious whether or not 
the CP violation by strong interactions of QCD (strong CP violation) certainly
exists and one could observe a finite size of its violation. 
To reveal the mystery, 
many theoretical and experimental challenges (such as detecting 
the electric dipole moment of neutron) have so far been met. 
However, the strong CP violation has not yet been detected 
because of its extremely tiny size.

The strong CP phase arises from the axial anomaly by 
converting the difference of the winding numbers ($\Delta N_W$),  
which is transferred by a topologically-nontrivial 
gluon configuration into  
the nonzero chirality ($\Delta N_{L-R}$); 
$(\Delta N_W \to \Delta N_{L-R})$. 
The difference of the winding numbers ($\Delta N_W$) 
is created by the transition 
between two of the infinite number of vacua   
forming the $\theta$ vacuum. 
In this sense, the detectability of the strong CP violation 
could depend on  
the survival probability of the transition rate with nonzero $\Delta N_W$. 
This transition rate 
is, however, known to be extremely smaller than the QCD time scale, 
which might thus cause difficulty in detecting the strong CP violation.

To access such  
detectability of the strong CP phase, 
it would be intriguing 
to place QCD in hot and/or dense environments. 
Because of the existence of 
the thermally excited-topologically nontrivial 
gauge configuration (the so-called  
sphaleron~\cite{Manton:1983nd,Klinkhamer:1984di}),  
in contrast to the vacuum case as above, 
the difference of winding numbers ($\Delta N_W$)  
can survive 
within the QCD time scale~\cite{McLerran:1990de,Moore:1997im,Moore:1999fs,Bodeker:1999gx}.   
When (almost massless) quarks couple to such a 
topological gauge configuration, 
the chirality imbalance ($\Delta N_{L-R}$)  
as well as  the strong CP phase would be present 
in the hot QCD system. 
The strong CP violation along with the chiral imbalance  
thus would (locally) be measured within the QCD time scale, 
due to the existence of the sphaleron configuration~\cite{McLerran:1990de,Moore:1997im,Moore:1999fs,Bodeker:1999gx}.

The existence of the chirality imbalance 
would imply the presence of 
a ``chemical potential'',  the so-called  {\it chiral chemical potential} $\mu_5$, 
which acts as a constant all over the hot QCD system 
and gives the energy difference $(\Delta {\cal E}_{L-R})$ between chiral quarks, 
such as $\Delta {\cal E}_{L-R}  = - \mu_5 \Delta N_{L-R}$~\cite{McLerran:1990de}. 
Such a local domain would therefore look like a medium, 
called the {\it chiral-imbalance medium}, where the preference of 
chiralities (such as $\Delta {\cal E}_{L-R}$) 
and the CP violation are locally present.

Heavy-ion collision experiments might form such metastable CP-odd 
domains characterized by the chiral chemical potential $\mu_5$, 
as has so far been addressed in the literature, e.g., 
\cite{Kharzeev:2001ev,Kharzeev:2007tn,Kharzeev:2007jp,Fukushima:2008xe} 
and \cite{Andrianov:2012hq,Andrianov:2012dj}.

In this paper, we discuss a novel possibility to observe the strong CP violation in 
the chiral-imbalance medium: the new probe might be  
a polarized photon tagged with a charged pion 
coming out of 
the chiral-imbalance medium created in heavy ion collisions.

\section{The chiral imbalance medium}

We begin with a brief review of the theoretical and 
phenomenological grounds.  
In heavy ion collision experiments, 
the hot QCD medium (the so-called fireball) 
is created at the time scale $\tau_{\rm fireball}$ 
of ${\cal O}(1{\rm -}10)$ fm after the collision 
(for a recent review, see \cite{deSouza:2015ena}). 
As discussed in the literature \cite{Kharzeev:2007tn,Kharzeev:2007jp,Fukushima:2008xe,Andrianov:2012hq,Andrianov:2012dj}, 
we may take the $\theta$ parameter 
to be position-time dependent, i.e., $\theta = \theta(x)$: 
the $\theta$ fluctuation in time will be interpreted as the chiral chemical potential 
$\mu_5$, 
as will be seen below. 
The QCD $\theta$ term then looks like 
\begin{equation} 
 \int_{\rm bulk} d^4 x \, \theta(x) \cdot \frac{g_s^2}{32 \pi^2} 
\epsilon^{\mu \nu \rho \sigma} {\rm tr} [G_{\mu\nu} G_{\rho\sigma} ] 
\,, \label{theta-term} 
\end{equation}
where the symbol ``bulk" denotes the volume of the fireball bulk  
and $G_{\mu \nu}$ is the gluon-field strength, $G_{\mu\nu} 
= \partial_\mu G_\nu - \partial_\nu G_\mu - i g_s [G_\mu, G_\nu]$ 
with the QCD gauge coupling $g_s$.

One should recall that the $\theta$ term is related to 
the axial-anomaly form for the isosinglet axial current $j_{5,\mu}$,  
\begin{eqnarray}
\partial^\mu j_{5,\mu}
=\sum _f 2m_f \bar q_f i \gamma_5 q_f  
- \frac{g_s^2 N_f}{16 \pi^2} 
\epsilon^{\mu \nu \rho \sigma} {\rm tr} [G_{\mu\nu} G_{\rho\sigma} ] 
\,, 
\label{axial-anomaly}
\end{eqnarray}
where $q_f$ denotes the quark field, $m_f$ is the quark mass, 
and $N_f$ stands for the number of the quark flavors. 
Using Eq.(\ref{axial-anomaly}), one finds that   
Eq.(\ref{theta-term}) goes like 
\begin{equation} 
 \int_{\rm bulk} d^4 x \, 
\left( \frac{\partial_\mu \theta(x)}{2N_f} \right) 
\cdot 
j^\mu_5(x) 
\,, \label{theta-term:2} 
\end{equation}
(up to the terms of the form $\theta \cdot m_f$ times pseudoscalar density) 
where 
integration by parts has been done. 
Looking at Eq.(\ref{theta-term:2}), 
we thus readily find that 
the time variation of $\theta$ plays the role of the 
chiral chemical potential~\cite{Kharzeev:2007tn,Kharzeev:2007jp,Fukushima:2008xe,Andrianov:2012hq,Andrianov:2012dj}: 
\begin{eqnarray} 
&& \left( \frac{\partial_0 \theta(x)}{2N_f} \right) \equiv \mu_5
\nonumber\\ 
{\rm s.t.} 
&& 
\int_{\rm bulk} d t \, 
\mu_5 \cdot 
Q_5(t)  = - \mu_5 \Delta N_{L-R} 
\,, 
\end{eqnarray}  
where $\Delta N_{L-R}= - \int dt Q_5(t) = - \int d^4x j_5^0(x)$. 
In that case the chiral chemical potential is 
embedded into the time component of the 
axialvector source ${\cal A}_\mu^0$ as~\cite{Fukushima:2008xe}   
\begin{equation} 
 {\cal A}_\mu^0 
= (\mu_5, \vec{0})^T 
\,. 
\end{equation} 
Through the axialvector source coupled to the quark chiral current, 
we are thus able to address hadron physics in the chiral-imbalance medium, 
where  
the size of $\mu_5$ is expected to be on the order of QCD scale, ${\cal O}(100\, 
{\rm MeV})$.

In the presence of the chiral chemical potential $\mu_5$, 
the parity symmetry as well as the Lorentz symmetry are  broken, 
and one is then left with the three-dimensional $SO(3)$ symmetry. 
This parity breaking effect would be transferred by hadrons, 
for instance, through their interactions with external gauge currents, 
arising 
from the intrinsic-parity odd form like the $\pi^0 \gamma\gamma$ 
vertex. (Here the intrinsic parity is defined to be even when 
a particle has the parity $(-1)^{\rm spin}$, otherwise it is odd.)  
Several studies regarding this kind of parity breaking effects 
have so far been done in the context of hadron physics under 
a strong field configuration (e.g., \cite{Fukushima:2012fg,Cao:2015cka}), 
or under the influence of the nonzero chiral chemical potential 
without strong fields~\cite{Andrianov:2012hq,Andrianov:2012dj}.

\section{Pion form factors in the chiral imbalance medium}

%In the present paper, 

We propose a new parity violating 
process related to the pion-electromagnetic form factors  
arising from intrinsic parity-odd interactions 
in the chiral imbalance medium. 
We derive its phenomenological consequence to discuss  
a new detection possibility for the strong CP violation 
in heavy ion collision experiments.

We first consider the general expression for the form factors of the charged pion 
coupled to the background photon field $A^\mu=(A^0, A^i=\vec{A})$ ($i=1,2,3$) 
in the presence of parity violation and $SO(3)$ symmetry.  
It is written as 
\begin{eqnarray}
&&
\langle\pi^+(p'=q+p)|{\cal J}^\mu(0) |\pi^+(p) \rangle 
\nonumber \\ 
&=&
e \, F_1 \cdot (p^0 + p'^0) \delta_0^\mu 
+ e \, F_2 \cdot(p^i + p'^i) \delta_i^\mu 
\nonumber\\
&& 
+ e \, F_3 \cdot(q^i) \delta_i^\mu 
+ ie \, F_4 \cdot \epsilon^{0 \mu \nu \rho} p_\nu p'_\rho  
\,, 
\label{general} 
\end{eqnarray}
where $q$ denotes the transfer momentum $q$;  
${\cal J}^\mu$ is the electromagnetic current (including 
nonminimal forms in general); $e$ is the electromagnetic coupling 
and 
$F_{i=1,2,3,4}=F_{i=1,2,3,4}(q^0,p^0,|\vec{q}|,|\vec{p}|)$ denote the form factors. 
In Eq.(\ref{general}), just for later convenience,   
we have defined the form factor $F_1$ by extracting the factor $(p^0+p'^0)$. 
In terms of the effective action, the first term and second term correspond to 
the $\pi^+-\pi^--\gamma$ interactions such as $A^0j^0$ and ${\vec A}\cdot {\vec j}$, respectively,  
where $j^\mu$ is the pion electromagnetic current of the minimal coupling form: 
$j_\mu=ie(\partial_\mu\pi^+\cdot \pi^--\partial_\mu\pi^-\cdot\pi^+)$. 
The third term can be absorbed into a redefinition of 
the other form factors by imposing the Ward-Takahashi identity 
for the electromagnetic gauge invariance. 
The fourth term $F_4$ is the one 
that will be induced in the presence of $\mu_5$ in the chiral-imbalance medium,  
as will be shown below~\footnote{
The $F_4$ type form factor would also be 
induced when one employs 
some dense matter system with 
other chiral chemical potentials including the isospin breaking effect   
(called the chiral-isospin chemical potential; e.g., see~\cite{Ebert:2016hkd}).  }.

To see how the $F_4$ form factor shows up  
in the chiral-imbalance medium with nonzero $\mu_5$, 
we employ 
the chiral Lagrangian based on the coset space $G/H=[U(2)_L\times U(2)_R]/U(2)_V$ 
and the Wess-Zumino-Witten (WZW) term~\cite{Wess:1971yu}: 
\begin{eqnarray} 
S=\int d^4 x \frac{f_\pi^2}{4}{\rm tr}[D_\mu U^\dagger D^\mu U ]
+ S_{\rm WZW} 
\,, \label{Lag}
\end{eqnarray}
where $f_\pi (\simeq 92 \,{\rm MeV})$ is the pion decay constant, 
and $U$ denotes the chiral field parameterizing the pion fields as 
$U=\exp( \sum_{a=1}^{3} 
2i\pi^a (\sigma^a/2)/f_\pi)$ with $\sigma^a$ 
being the Pauli matrices. 
Here we simply ignored the $U(1)_A$ pion component because it is irrelevant to 
the present our proposal. 
The covariant derivative acting on $U$ is defined as    
$D_\mu U=\partial_\mu U-i{\cal L}_\mu U+iU{\cal R}_\mu$ with the external gauge fields ${\cal L}_\mu$ and ${\cal R}_\mu$ 
arising by gauging the chiral $U(2)_L \times U(2)_R$ symmetry.  
The photon field and the chiral chemical potential $\mu_5$ are then 
introduced as a part of the external vector and axialvector gauge fields 
$({\cal V}_\mu, {\cal A}_\mu)$ as  
\begin{eqnarray} 
{\cal V}_\mu&=&\frac{1}{2}({\cal R}_\mu+{\cal L}_\mu)=eQ_{\rm em}A_\mu
\,, \nonumber\\
{\cal A}_\mu&=&\frac{1}{2}({\cal R}_\mu-{\cal L}_\mu)=\mu_5 \, \delta_{\mu0} \cdot 1_{2\times2},
\end{eqnarray}
where the electromagnetic charge matrix is defined as $Q_{\rm em}=\sigma_3/2 + 1_{2\times 2}/6$.

The WZW term 
in Eq.(\ref{Lag}) includes the following interactions between 
the pion and external gauge fields: 
\begin{eqnarray} 
S_{\rm WZW}&=&
i\frac{N_c}{48\pi^2} 
\int d^4 x \, \epsilon^{\mu\nu\rho\sigma}{\rm tr}\Bigl[\{\partial_\mu {\cal L}_\nu,{\cal L}_\rho\}(\partial_\sigma U)U^{\dag}
\nonumber\\
&&+\{\partial_\mu{\cal R}_\nu,{\cal R}_\rho \}U^{\dag}\partial_\sigma U
-\partial_\mu{\cal L}_\nu\partial_\rho U{\cal R}_\sigma U^{\dag}\nonumber\\
&&+\partial_\mu{\cal R}_\nu\partial_\rho U^{\dag}{\cal L}_\sigma U-{\cal R}_\mu U^{\dag}{\cal L}_\nu(\partial_\rho U)U^{\dag}\partial_\sigma U\nonumber\\
&&+U^{\dag}{\cal L}_\mu U{\cal R}_\nu U^{\dag}(\partial_\rho U)U^{\dag}(\partial_\sigma U)\nonumber\\
&&
-\frac{1}{2}{\cal L}_\mu(\partial_\nu U)U^{\dag}{\cal L}_\rho(\partial_\sigma U)U^{\dag}\nonumber\\
&&+\frac{1}{2}{\cal R}_\mu (\partial_\nu U)U^{\dag}{\cal R}_\rho(\partial_\sigma U)U^{\dag}\Bigl]
+ \cdots 
\,,  
\end{eqnarray}
with the number of colors $N_c=3$. 
Thus, we extract the electromagnetic interactions of the charged pion in the chiral-imbalance medium 
including the WZW term to get 
\begin{eqnarray} 
D_\mu\pi^+ D^\mu \pi^-+\frac{ie\mu_5N_c}{6\pi^2f_\pi^2} \epsilon^{0\nu\rho\sigma}\partial_\nu\pi^+\partial_\rho\pi^- A_\sigma
\,, \label{Lag:2}
\end{eqnarray}
where $D_\mu\pi^\pm = (\partial_\mu \mp ie A_\mu)\pi^\pm$. 
Obviously, the second term breaks the parity, but preserves the charge conjugation, 
which
indicates the CP violation with nonzero $\mu_5$. 
From Eq.(\ref{Lag:2}) we read off the electromagnetic form factors corresponding to 
the general expressions in Eq.(\ref{general}):  
\begin{eqnarray} 
&&F_1=F_2=1 
\,, \nonumber\\
&&F_3=0 
\,, \nonumber\\
&&F_4=\frac{\mu_5 N_c}{6\pi^2 f_\pi^2} 
\,. 
\label{form-factors}
\end{eqnarray}

The result on $F_1=1$ in Eq.(\ref{form-factors}) 
just tells us that 
the pion electromagnetic charge is 
correctly normalized, which just 
reflects the electromagnetic gauge invariance, 
while the normalization for $F_2=1$ has happened by accident.   
It is interesting to note that, 
although the pion is manifestly the spin 0 particle, 
the nonzero $F_4$ term looks like 
the anomalous magnetic moment of the charged pion like 
${\vec B} \cdot {\vec j}$, when the charged pions are 
set in an oriented-magnetic field background.

%\begin{widetext}
\begin{figure}[t] 
\vspace{-1.5cm}
\begin{center}
\includegraphics*[scale=0.4]{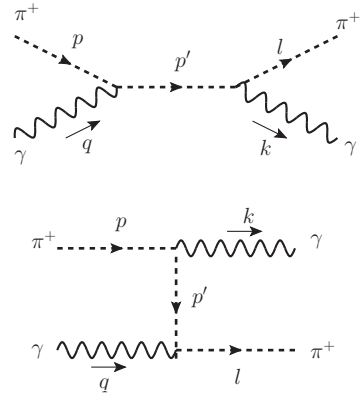}
\vspace{-2cm}
\caption{ 
The Feynman graphs illustrating 
$\pi^\pm \gamma  \to \pi^{\pm} \gamma$ scattering process 
involving the parity breaking form factor $F_4$ 
and the parity conserving $F_1(=F_2)$ 
in Eq.(\ref{form-factors}).   
} 
\label{fig1}
\end{center}

\end{figure}
%\end{widetext}

\section{Measuring parity violation in the pion-photon scattering }

Now we propose a new possibility to detect the remnant of 
the strong CP violation mimicked by nonzero chiral chemical potential 
$\mu_5$ involved in the $F_4$ form factor.  
The $\mu_5$-dependent $F_4$ generates 
$\pi^\pm \gamma  \to \pi^{\pm} \gamma$ scattering processes as 
depicted in Fig.~\ref{fig1}~\footnote{  
Actually, one needs another contribution arising from a contact 
$\pi^\pm \gamma \pi^\mp \gamma$ vertex present in Eq.(\ref{Lag}),   
to make the scattering amplitude gauge invariant.}.    
In light of the heavy ion collisions, 
the charged pion and photon in the initial state (corresponding to 
the left-side states for each graph in Fig.~\ref{fig1}) 
come from the chiral-imbalance medium, 
so we consider the unpolarized photon with 
the sum and average over spin properly taken. 
In the final state (depicted in the right-side states for each graph in Fig.~\ref{fig1}) 
we impose that the photon be polarized with certain spin $(s)$ so as to 
make the helicity dependence $(\lambda \propto \vec{k} \cdot \vec{s})$ 
with respect to the photon outgoing momentum ($\vec{k}$). 
Thus the following two processes are parity conjugate:  
\begin{eqnarray} 
\pi^\pm(\vec{p}) + \gamma(\vec{q})  
&\to &
 \pi^\pm(\vec{l}) + \gamma_{+}(\vec{k})   
 \,, \nonumber \\ 
 \pi^\pm(-\vec{p}) + \gamma(-\vec{q})  
 &\to& 
 \pi^\pm(-\vec{l}) + \gamma_{-} (-\vec{k})  
 \,,   \label{pi-gamma-process}
\end{eqnarray}  
where $\pm$ attached on photons in the final state denote the photon helicities.

\begin{figure}[t] 
\centering
\includegraphics*[scale=0.4]{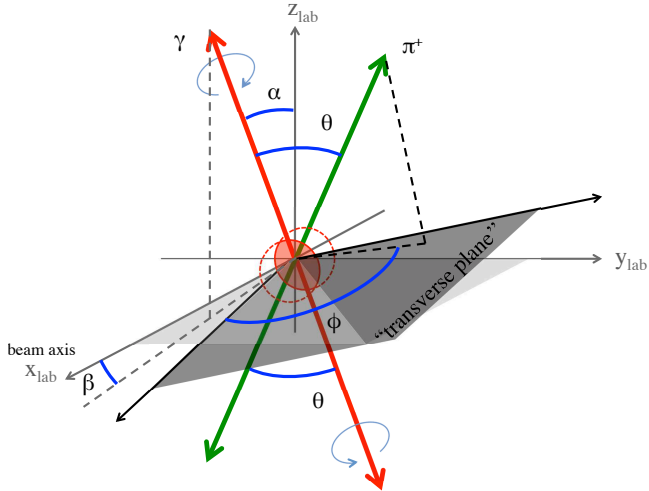}
\caption{ 
A schematic sketch of the laboratory frame geometry 
relevant to the proposed 
parity violation 
measurement in heavy ion collisions. 
The laboratory frame is spanned by the coordinates 
$(x_{\rm lab}, y_{\rm lab}, z_{\rm lab})$. 
The red blob at the origin denotes the fireball domain created by 
the collision of heavy ions (red dashed circles).  
The plane drawn in blue is set to 
be transverse to the photon direction, which we call the ``transverse plane''.    
The circles with arrows surrounding around the photon profile (red allows) 
represent the helicities. 
The pion trajectories (displayed as green profiles) are specified  
by the azimuthal angle $\phi$ measured in the ``transverse plane''  
and the zenith angle $\theta$ 
with respect to the photon profiles  
determined by the angles $\alpha$ and $\beta$ 
in the laboratory frame. 
}
\label{fig2}
\end{figure}

The geometrical configuration relevant to the processes 
in Eq.(\ref{pi-gamma-process}) 
in heavy ion collision experiments 
is sketched in Fig.~\ref{fig2}. 
In the laboratory frame (defined at the rest frame of the observer/detector),  we measure 
charged pions and photons emitted from the fireball bulk,   
%In the lab frame 
where the photons are specified by 
the zenith angle $\alpha$ and the azimuthal angle $\beta$ 
in the laboratory coordinate space $(x_{\rm lab}, y_{\rm lab}, z_{\rm lab})$. 
Then the location of the charged pion is determined 
by the opening angle $\theta$ from the photon 
and 
another azimuthal angle $\phi$  
on the plane transverse to 
the photon direction, which we shall call the ``transverse plane'' 
(see Fig.~\ref{fig2}). 
The angles $\theta$ and $\phi$ correspond to
 kinematical variables in the final state in computing
 the $\pi^\pm \gamma$ scattering amplitudes.  
As the parity conjugate event, we also observe 
the mirror process at the laboratory frame position 
with the same opening angle  
$\theta$ as well as the azimuthal angle $\phi$
between the charged pion and photon.

We shall suppose that those photon polarizations could be observed, say, by 
the circular polarization apparatus designated in the laboratory frame, 
with the angles $\alpha$ and $\beta$ taken into account. 
In that case, the number of these mirror events could be 
different due to the parity violation induced by the $\mu_5$-dependent 
$F_4$ term. 
Such an asymmetry $({\cal A})$ can be evaluated as  
\begin{eqnarray}
{\cal A} 
= 
\Bigg| 
\frac{ {\cal N}_+   -  {\cal P} [{\cal N}_+] }
{ \sum_\lambda 
\{ 
{\cal N}_\lambda 
+ 
{\cal P}[{\cal N}_\lambda]  \}} \Bigg|  
\,, \label{A:def}
\end{eqnarray} 
where ${\cal N}_\lambda$ stands for 
the number of events per the phase space, $dE_\gamma d\cos\theta d\phi$,  
for the parity 
conjugate processes in Eq.(\ref{pi-gamma-process}) with the helicity $\lambda$ 
and the photon energy $E_\gamma$  
in the final state. 
The symbol ${\cal P}$ acts as the parity conjugation projection. 
The denominator 
represents the total number of the $\pi^\pm \gamma$ emission 
events with unpolarized photons per the phase space.

One may notice that the asymmetry in Eq.(\ref{A:def}) 
generically depends on the kinematical variables 
associated with the initial state particles in the medium.  
Since the chiral imbalance medium is formed in the hot QCD bulk, 
it would be reasonable to take 
the thermal average over the phase space for the initial state,  
which would be subject to the strong dependence on  
modeling of the hot QCD medium.

Instead of the detailed computation including the thermal 
effects and full channel analysis contributing to the $\pi^\pm \gamma$ 
scattering processes,  
here we may just estimate the typical size of 
the asymmetry, by focusing only on the pion resonant channel 
(the upper graph in Fig.~\ref{fig1}), 
to get a rough implication of our proposal. 
In this case, the asymmetry would be free from  
kinematical configuration in the initial state and 
only depend on  
the final state variables ($\theta, \phi, E_\gamma$ and the pion energy $E_\pi$), 
and hence we are 
allowed to express the asymmetry in Eq.(\ref{A:def}) as 
\begin{eqnarray}
&& {\cal A}^{\rm s-channel}(E_\gamma, E_\pi, \theta, \phi) 
= 
\nonumber \\ 
&& 
\Bigg| \frac{ {\cal N}_+ (E_\pi, E_\gamma, \theta,\phi)  
-  {\cal P} [{\cal N}_+(E_\pi, E_\gamma, \theta, \phi)] }{ \sum_\lambda 
\{ {\cal N}_\lambda(E_\pi, E_\gamma, \theta,\phi) 
+ 
{\cal P} [ {\cal N}_\lambda (E_\pi, E_\gamma,\theta, \phi) ] \}  } \Bigg|  
\,. 
\nonumber \\ 
\label{A-s:def} 
\end{eqnarray} 
 Using Eq.(\ref{A-s:def})
 we evaluate the typical size of the asymmetry by 
 choosing  the circular polarization state for the photon 
 in the final state. 
Note that the asymmetry arises from the $\mu_5$ term along with the loop suppression 
factor as
\begin{equation} 
\frac{N_c \mu_5 E_{\pi/\gamma}}{(4 \pi f_\pi)^2} 
\sim 
\frac{\mu_5}{ (4\pi f_\pi)} ={\cal O}(10^{-1})
\,,  
\end{equation} 
with $E_{\pi/\gamma} = {\cal O}(4 \pi f_\pi) 
= {\cal O}(1)$ GeV 
and 
$\mu_5 = {\cal O}(100)$ MeV being assumed.  
We may therefore expand  
terms in Eq.(\ref{A-s:def}) 
in powers of that loop factor and keep only the nontrivial 
leading order contributions 
to get  
\begin{eqnarray}
&& {\cal A}^{\rm s-channel} 
= \nonumber \\ 
&& 
 \frac{\mu_5 N_c}{12 \pi^2 f_\pi^2} 
 \frac{E_\gamma (E_\pi^2 - m_\pi^2) \sin^2\theta}{
 m_\pi^2 + E_\gamma E_\pi \left(1 - \sqrt{1 - \frac{m_\pi^2}{E_\pi^2}} \cos\theta  \right)
} 
\,.  \label{A:calc} 
\end{eqnarray}  
In the limit where $E_{\pi/\gamma} \gg m_\pi$ and $\theta \to 0$,  
we find that the asymmetry reaches the maximum:  
\begin{eqnarray}  
{\cal A}^{\rm s-channel}|_{\rm max}  
&=& \frac{\mu_5 E_\pi N_c}{6 \pi^2 f_\pi^2} 
\nonumber \\ 
&\simeq&  
0.2 \times \left( \frac{\mu_5}{200\,{\rm MeV}} \right) 
\left(\frac{E_\pi}{1 \,{\rm GeV}}  \right) 
\,. \label{A:max}
\end{eqnarray} 
The estimated size of the asymmetry is indeed on the order of 
the loop suppression factor $(\mu_5/(4\pi f_\pi)) (={\cal O}(10^{-1}))$ 
as we expected.

Possible background events would come from 
the electroweak interactions breaking the parity, 
for instance, from the W boson exchange process 
in the vacuum through the $W$-$\pi$-$\gamma$ 
vertex arising in the WZW term. 
As long as one focuses on the pion pole part as evaluated in the above,    
however, such a W boson contribution would negligibly be small  
compared with the pion exchange term, roughly by a factor of 
${\cal O}(G_F^2 f_\pi^4) = {\cal O}(10^{-14})$, 
where $G_F$ is the Fermi constant 
$\simeq (290\,{\rm GeV})^{-2}$ and the  
typical order of all the 
transferred momenta 
and $\mu_5$ are set to $f_\pi$. 
Thus the proposed asymmetry arises dominantly from the parity violation by  
 $\mu_5 \neq 0$, namely, the strong CP violation, 
which could be testable in heavy ion collisions with high statistics.

\section{Conclusion}

In this paper, 
we have discussed a possibility for charged pions to 
act as a probe for measuring the strong CP violation: 
the charged-pion along with the photon 
could pick up the remnant of the CP violation 
in the chiral-imbalance medium, which can be 
produced in heavy ion collisions. 
Our proposal presented here could be tested 
through the parity violation in a lab frame (defined in the text),  
by measuring 
the polarized photons tagged by 
charged pions with finite opening angle in the lab frame of 
heavy ion collision experiments.

We have roughly estimated the typical size of 
the parity violation, 
quantified as the asymmetry, 
by focusing only on the pion exchange contribution 
in the so-called s-channel 
and find that 
the signal is much larger than the size of 
expected background from electroweak process.  
Inclusion of other channel contributions 
would not significantly make a difference from 
our estimate as far as the order of magnitude is concerned.

One also notices that our estimate was made by assuming 
the energy of charged pion to be much larger than the pion mass 
($E_\pi ={\cal O}(1)$ GeV). 
One may therefore naively suspect that other higher resonances such as 
the rho meson would not be negligible. 
However, 
it turns out that 
such higher resonance ($R$) contributions are 
necessarily suppressed by a factor of $(m_\pi^2/m_R^2) \ll 1$
in the asymmetry,  
when the nearly collinear emission $(\theta \simeq 0)$ 
is selected, as done in Eq.(\ref{A:max}): 
the propagator of the resonance $R$ with mass $m_R (\gg m_\pi)$ 
in the s-channel 
takes the form in the laboratory frame  
$\sim [E_\pi E_\gamma (1 - \sqrt{1-m_\pi^2/E_\pi^2} \cos\theta)- (m_R^2 - m_\pi^2)]^{-1}$. 
When the charged pion is emitted almost collinear to the photon (i.e. $\theta \simeq 0$) 
with the high energies $E_{\gamma/\pi} \gg m_\pi$,  
the $R$-propagator goes like  
$\sim m_R^{-2}$. 
Thus the estimated size of the asymmetry in Eq.(\ref{A:max}) 
will be reliable even for the high energy emission of charged 
pions.

Note that our parity violation induced by nonzero chiral chemical 
potential can be seen in the laboratory frame, but 
cannot be seen in the resonance-mass distribution in the usual 
center of mass frame, such as in the dilepton mass distribution, 
as discussed in the literature~\cite{Andrianov:2012hq}.  
In this sense, our findings presented in this paper 
certainly include a new proposal to probe the strong CP 
violation in a hot QCD medium, and could give 
a large enough signal in a realistic experimental situation 
for heavy ion collisions.

More precise estimates of the charged pion-photon scattering 
amplitudes including the thermal effect in the medium and   
the chiral-imbalance effects involving other light hadrons 
will be studied elsewhere, which would make our prediction more 
definite.

In addition to the theoretical uncertainties, 
however,  one needs to more seriously take into account 
the experimental setup in heavy ion collisions.  
Since thousands of particles are produced in heavy-ion
collisions, the analysis of specifically selected processes, 
such as elastic scattering of pions and photons proposed in the present 
paper, would be badly contaminated by the
background effects; thermal emission, inelastic scattering, and so forth. 
Due to those complex experimental configurations, 
the energy range for the polarization
asymmetry would need in particular to be limited to 
relatively low photon energy ranges, 
where the background would still get too large to be simply understood. 
Though being  the definite signal for the strong CP violation,  
the presently proposed parity violation measurement 
would therefore be very challenging and would
require high statistics and resolution, e.g., for the photon energy.

In closing, we shall give some comments on 
other possible phenomenological implications derived from 
the $\mu_5$-dependent form factor.

Besides the parity violation as noted above, 
actually,  
our $\mu_5$-induced form factor $F_4$ would 
give rise to other heavy-ion collision signatures,    
much like the so-called the chiral magnetic effect~\cite{Kharzeev:2007tn,Kharzeev:2007jp,Fukushima:2008xe}, 
which can be seen if an oriented-magnetic field 
is present in the fireball bulk.   
In particular, the $F_4$ form factor in Eq.(\ref{form-factors}) generates 
the $\vec{B} \cdot \vec{j}$ in the magnetic field background,  
hence the $F_4$ contribution would contaminate the event-by-event charge separation 
measurements signaling the chiral magnetic effect~\cite{Abelev:2009ac,Voloshin:2010ju}.  
In addition, our direct photon without the specified polarization 
%and tagging charged pions   
could also mix with the signal generated from 
the chiral magnetic effect as discussed in~\cite{Fukushima:2012fg}.

\acknowledgments 

We are grateful to Kenji Fukushima and Kazunori Itakura for 
useful comments. 
This work was supported in part by 
the JSPS Grant-in-Aid for Young Scientists (B) No. 15K17645 (S.M.), 
and the JSPS Grant-in-Aid for Scientific Research (C)
No. 16K05345 (M.H.). 
R.O. was partially supported by the 
TAQ honor program in physics from the Office of Undergraduate 
Education and College of Physics in Jilin University.

%%%%%%%%%%%%%%%%%%%%%%%%%%%%%%%%%%%%%%%%%%%%%%%%%%%%%%%%%%%%%%%%%%%%%%%%%%%%%%%%%%%%%%%%%%%

\end{document}